\documentclass[journal]{IEEEtran}
\usepackage{amsmath,graphicx,url,times}
\usepackage{amsfonts}
\usepackage{booktabs}
\usepackage{algorithm}
\usepackage[noend]{algpseudocode}
\usepackage[T1]{fontenc}
\newcommand{\qk}[1] {{\color{black}#1}}

\usepackage[usenames, dvipsnames]{color}
\usepackage{booktabs}
\usepackage{multirow}
\ifCLASSINFOpdf
\else
\fi
\begin{document}
%
% paper title
% Titles are generally capitalized except for words such as a, an, and, as,
% at, but, by, for, in, nor, of, on, or, the, to and up, which are usually
% not capitalized unless they are the first or last word of the title.
% Linebreaks \\ can be used within to get better formatting as desired.
% Do not put math or special symbols in the title.
\title{{\huge Sound Event Detection of Weakly Labelled Data with CNN-Transformer and Automatic Threshold Optimization}}
%
%
% author names and IEEE memberships
% note positions of commas and nonbreaking spaces ( ~ ) LaTeX will not break
% a structure at a ~ so this keeps an author's name from being broken across
% two lines.
% use \thanks{} to gain access to the first footnote area
% a separate \thanks must be used for each paragraph as LaTeX2e's \thanks
% was not built to handle multiple paragraphs
%

%\author{Michael~Shell,~\IEEEmembership{Member,~IEEE,}
%        John~Doe,~\IEEEmembership{Fellow,~OSA,}
%        and~Jane~Doe,~\IEEEmembership{Life~Fellow,~IEEE}% <-this % stops a space
%\thanks{M. Shell was with the Department
%of Electrical and Computer Engineering, Georgia Institute of Technology, %Atlanta,
%GA, 30332 USA e-mail: (see http://www.michaelshell.org/contact.html).}% <-this % stops a space
%\thanks{J. Doe and J. Doe are with Anonymous University.}% <-this % stops a space
%\thanks{Manuscript received April 19, 2005; revised August 26, 2015.}}

\author{Qiuqiang Kong*,~\IEEEmembership{Student Member,~IEEE}, Yong Xu*,~\IEEEmembership{Member,~IEEE}, \\ Wenwu Wang,~\IEEEmembership{Senior Member,~IEEE} and Mark D. Plumbley,~\IEEEmembership{Fellow,~IEEE}% <-this % stops a space
\thanks{This work was supported in part by the EPSRC Grant EP/N014111/1 ``Making Sense of Sounds'', in part by the Research Scholarship from the China Scholarship Council 201406150082, and in part by a studentship (Reference: 1976218) from the EPSRC Doctoral Training Partnership under Grant EP/N509772/1.}\thanks{Q. Kong, and M. D. Plumbley are with the Centre for Vision, Speech and Signal Processing, University of Surrey, Guildford GU2 7XH, U.K. (e-mail: q.kong@surrey.ac.uk; m.plumbley@surrey.ac.uk).}
\thanks{Y. Xu is with the Tencent AI Lab, Bellevue, WA 98004 USA (e-mail:
lucayongxu@tencent.com).}
\thanks{W. Wang is with the Centre for Vision, Speech and Signal Processing,
University of Surrey, Guildford GU2 7XH, U.K., and also with Qingdao
University of Science and Technology, Qingdao 266071, China (e-mail:
w.wang@surrey.ac.uk).}}

\maketitle

% As a general rule, do not put math, special symbols or citations
% in the abstract or keywords.
\begin{abstract}
Sound event detection (SED) is a task to detect sound events in an audio recording. One challenge of the SED task is that many datasets such as the Detection and Classification of Acoustic Scenes and Events (DCASE) datasets are weakly labelled. That is, there are only audio tags for each audio clip without the onset and offset times of sound events. \qk{We compare segment-wise and clip-wise training for SED that is lacking in previous works. We propose a convolutional neural network transformer (CNN-Transfomer) for audio tagging and SED, and show that CNN-Transformer performs similarly to a convolutional recurrent neural network (CRNN)}. Another challenge of SED is that thresholds are required for detecting sound events. Previous works set thresholds empirically, and are not an optimal approaches. To solve this problem, we propose an automatic threshold optimization method. The first stage is to optimize the system with respect to metrics that do not depend on thresholds, such as mean average precision (mAP). The second stage is to optimize the thresholds with respect to metrics that depends on those thresholds. \qk{Our proposed automatic threshold optimization system achieves a state-of-the-art audio tagging F1 of 0.646, outperforming that without threshold optimization of 0.629, and a sound event detection F1 of 0.584, outperforming that without threshold optimization of 0.564.}

\end{abstract}

% Note that keywords are not normally used for peerreview papers.

\IEEEpeerreviewmaketitle

\section{Introduction}

Sound event detection (SED) is an important research topic which can be used in smart home, self-driving cars and smart cities. For example, a SED system can detect an ambulance siren even if the ambulance is far away. In this situation, it is difficult to detect the siren with cameras because of the distance and obstructions. Different from audio tagging (AT) which only requires to detect the presence or absence of sound events in an audio recording, SED requires to predict the onsets and offsets of sound events. SED has attracted many researches since the introduction of the Detection and Classification of Acoustic Scenes and Events (DCASE) challenges \cite{giannoulis2013detection, stowell2015detection, mesaros2016tut, mesaros2017dcase, mesaros2018multi}.

One challenge of the SED task is that audio recordings are usually weakly labelled. That is, in the training data, we only know the presence or absence of sound events, without knowing their onset and offset times. We call this kind of data \textit{weakly labelled data (WLD)}. In this paper, we focus on the large-scale weakly supervised sound event detection task for smart cars dataset from the DCASE 2017 challenge Task 4 \cite{dcase2017task4}. The audio recordings from this task is a subset of the AudioSet dataset \cite{audioset}. This task includes both AT and SED. All audio clips for training are weakly labelled without time information of sound events. In previous research of AT, several CNN-based methods have been applied \cite{cakir2016domestic, yongIJCNN2017, hershey2017cnn, kumar2016audio, xu2017large, su2017weakly, chou2018learning, Salamon2017, Lee2017a, Chou2017, adavanne2017sound, ford2019deep}. Those approaches show that a robust feature extractor is important for AT and SED. Usually CNNs are applied to the log mel spectrogram of audio recordings followed by a sigmoid non-linearity to predict the presence probabilities of sound events. 

General SED tasks can be divided into two categories according to the availability of frame-level or clip-level labels: strongly supervised SED, when frame-level labels are provided; or weakly supervised SED, when only clip-level tags are provided. Several deep learning based methods \cite{cakir2015polyphonic, parascandolo2017convolutional, hayashi2017blstm, dai2016sound, xia2017frame, phan2016robust, wang2017first, gururani2018instrument} have been proposed for the strongly supervised SED task. However, frame-level sound event labels are time consuming to obtain. Recently, the DCASE 2017 Task 4 provides a large-scale dataset designed for AT and SED with only weakly labelled data provided. \qk{To train with weakly labelled data, segment-wise based methods \cite{hershey2017cnn, Lee2017} split audio clips into segments, and assign each segment with weak labels. On the other hand, clip-wise training methods \cite{xu2017large} apply entire audio clips for training. There is a lack of research comparing the segment-wise and clip-wise based methods for SED.}

Although previous CNNs based methods have been successful in audio tagging and SED tasks, CNNs do not capture the long time dependency in an audio clip well. For example, the receptive field of a CNN can be limited to a short duration with a fixed length that does not take long history information into account in the system. \qk{To solve this problem, convolutional recurrent neural networks (CRNNs) including \cite{choi2017convolutional, cakir2017convolutional, adavanne2017sound} and bidirectional long short term memory (BLSTM) systems \cite{hayashi2017blstm} were used to consider the long temporal information for audio tagging and sound event detection}. One disadvantage of CRNNs is that the hidden states of a CRNN have to be calculated one by one, and can not be calculated in parallel. Recently, transformers \cite{vaswani2017attention, devlin2018bert, won2019toward} have been proposed to consider the long time dependency of sequences. A transformer consists of several attention layers. Each state of a layer takes the information from all states of the previous layer. Therefore, each state retains the global information of the input sequence.  

Another challenging problem of AT and SED is the selection of thresholds for post-processing \cite{gururani2018instrument, cances2019evaluation}. For example, in the AT subtask, if the predicted probability of a sound class is over a threshold in an audio clip, then the audio clip is regarded as containing this sound class. The thresholds selection is an important part of AT and SED. Usually, the thresholds are selected empirically. For example, in the winning system of the AT subtask in the DCASE 2017 \cite{xu2017large}, thresholds of 0.3 are used for all the sound classes. However, those thresholds are selected by experience and may not be optimal. In this work, we propose an automatic threshold optimization method to solve this problem. 

This work contributes in the following aspects. First, we investigate segment-wise training and clip-wise training for AT and SED. We found that different systems perform differently for the AT and SED subtask. Second, we propose a CNN-Transformer system, and achieves competitive results to the CNN-GRU system. Third, we propose an automatic threshold optimization method for the AT and SED subtask. Our proposed systems outperform the best systems in the DCASE 2017 Task 4 challenge. This paper is organized as follows: Section \ref{section:cnn} introduces CNN and CRNN for AT and SED. Section \ref{section:cnn_transformer} introduces a CNN-Transformer system. Section \ref{section:segment_clip_wise} introduces segment-wise and clip-wise training. Section \ref{section:thresholding} proposes an automatic threshold optimization method for AT and SED. Section VI shows experimental results. Section \ref{section:conclusion} concludes this work.

%demo file is intended to serve as a ``starter file''
%for IEEE journal papers produced under \LaTeX\ using
%IEEEtran.cls version 1.8b and later.
% You must have at least 2 lines in the paragraph with the drop letter
% (should never be an issue)
%I wish you the best of success.

%\hfill mds
 
%\hfill August 26, 2015
\section{Convolutional neural networks (CNNs)}\label{section:cnn}
\subsection{Conventional CNNs}
CNNs were originally designed for image classification \cite{krizhevsky2012imagenet}, and have been recently used for audio related tasks such as speech recognition \cite{abdel2014convolutional} and AT \cite{choi2016automatic, kong2018dcase}. A conventional CNN consists of convolution layers, pooling layers and fully connected layers. The input to each convolutional layer is a tensor with a shape $ (N, C, W, H) $ representing the number of input samples, channels, width and height. For AT, the input width and height represent the number of time frames and frequency bins. Each convolutional layer consists of a set of learnable kernels. The output of a convolutional layer is a tensor called feature maps. The kernels in a convolutional layer can learn local time-frequency patterns in the spectrogram of an audio clip. \qk{In audio processing, low level features \cite{thickstun2016learning} can be waveforms or time-frequency representations such as spectrogram. High level features are those extracted from low level features by convolutional layers}. Recent CNN architectures apply batch normalization \cite{ioffe2015batch} after convolutional layers to speed up and stabilise training. Nonlinear activation functions such as ReLU \cite{nair2010rectified} are applied after each batch normalization. For AT and SED, pooling layers are applied along both time and frequency axes. A time distributed fully connected layer is applied on the output of the last convolutional layer to predict the presence probability of sound events along the time axis. Then the predicted probabilities are aggregated over the time axis to obtain the clip-wise sound event presence probability. The aggregation can be, for example, maximum or average operations over the time axis. 

\subsection{Convolutional recurrent neural network (CRNNs)}
The receptive field of CNNs have limited sizes. That is, CNNs can not capture long time dependency in an audio clip. However, some sound events have long time dependencies. For example, an ambulance siren may last for tens of seconds, and the temporal information is useful for AT and SED. Designing a system that is able to capture the temporal dependency is beneficial for AT and SED. Recurrent neural networks (RNNs) \cite{mikolov2010recurrent} are kinds of neural networks that can store history information in their hidden states, and thus capture long term dependency of sequential data. RNNs have been applied to language processing tasks such as \cite{mikolov2010recurrent}. The potential problem of a conventional RNN is that the gradient of weights may vanish or explode in training. Long short term memory (LSTM) \cite{hochreiter1997long} is a variation of RNN that introduces constant error carousel units, input gate, output gate and forget gate to avoid the gradient exploding and vanishing problem. An improved architecture of LSTM called gated recurrent units (GRU) \cite{cho2014learning} is proposed to reduce the parameters of LSTMs and simplify the gates to a reset gate and a forget gate. A GRU can be in both directions which we call bidirectional GRU (biGRU), which is applied in our AT and SED systems. 

\section{CNN-Transformer}\label{section:cnn_transformer}
CRNN can capture long time-dependency of sound events. On the other hand, the sequential nature of CRNNs also makes it more difficult to take advantage of modern fast computing devices such as GPUs. Recently, transformer \cite{vaswani2017attention} is proposed to learn correlations of time steps in a sequence such as natural language processing tasks \cite{devlin2018bert}. \qk{Compared with RNNs which require to compute the hidden states in a sequence, a transformer can parallelize the computation which only requires matrix multiplication in the forward pass.} Transformer applies a self-attention mechanism which directly models relationships between all time steps in a sequence. In an audio clip, a sound class may contain several sound events over time. For example, the speech of a human may appear in any time in an audio clip. A transformer can capture the correlation of speeches appearing in different part of an audio clip.

\subsection{Transformer}
Transformer was originally proposed in \cite{vaswani2017attention}. The motivation for the design of the transformer is to allow  modeling of dependencies without regard to their distance in the input sequence. In addition, a transformer allows for more parallel computing than RNNs by removing the recurrent connections. A transformer consists of several encoder and decoder layers. The encoder transforms an input to a high level embedding, and the decoder transforms an embedding to output. In a classification task such as AT or SED, we only need the encoder. Each encoder consists of several encoder layers. For each encoder layer, we denote the input to the encoder layer as a tensor $ x $ with a shape of $ T \times C $, where $T$ and $C$ represent the number of time steps and channels. We follow the symbols used in \cite{vaswani2017attention}. An encoder layer consists of a query transform matrix $ W^{Q} $, a key transform matrix $ W^{K} $ and a value transform matrix $ W^{V} $. The matrices $ W^{Q} $ and $ W^{K} $ have a shape of $ C \times d_{k} $, and $ W^{V} $ has a shape of $ C \times d_{v} $ where $ d_{k} $ and $ d_{v} $ are integers. Then the query $ Q $, key $ K $ and value $ V $ can be obtained by: 

\begin{equation}
\begin{split}
& Q = x W^{Q} \\
& K = x W^{K} \\
& V = x W^{V}. \\
\end{split}
\label{eq:qkv}
\end{equation}

\noindent The query $ Q $ and key $ K $ have a shape of $ T \times d_{k} $, and the value $ V $ has a shape of $ T \times d_{v} $. The output of an encoder layer can be written as: 

\begin{equation}
\begin{split}
& h = \text{softmax}(\frac{QK^{T}}{\sqrt{d_{k}}}) V,
\end{split}
\label{eq:transformer}
\end{equation}

\noindent where the output $ h $ has a shape of $ T \times H $. Equation (\ref{eq:transformer}) computes the dot product of the query with all keys, divide each by $ \sqrt{d_{k}} $, and apply a softmax function to obtain the weights on the values $ V $ \cite{vaswani2017attention}. The division of square root of $ d_{k} $ is a normalization term \cite{vaswani2017attention}. In (\ref{eq:transformer}), the inner product of $ Q $ and $ K^{T} $ has a shape of $ T \times T $, representing the feature correlation of different time steps. The softmax operation converts the correlation value to probabilities along the time steps indicating how much the value $ V $ in a time step should be attended.

\subsection{CNN-Transformer}
For audio tagging and SED, the input is usually a time-frequency representation such as a log mel spectrogram. Log mel spectrogram is a low level feature and CNNs have been proposed to apply on the log mel spectrogram to extract high level features \cite{choi2016automatic}. To build the CNN-Transformer system, we first apply a CNN described in Section \ref{section:cnn} on the log mel spectrogram of an audio clip. Convolutional layers in the CNN are used to extract high level features of the input log mel spectrogram. We use the feature maps of the last convolutional layer to obtain embedding vectors along time axis. The embedding can be viewed as $ x $ with a shape of the number of time frames by the number of channels. The output of the transformer has a shape of $ T \times d_{v} $. A fully connected layer followed by a sigmoid non-linearity is applied on this output to predict the presence probabilities of sound classes over time steps. An aggregation function such as average aggregation can be applied to average out those probabilities along time domain to obtain the audio tagging result. 

\section{Segment-wise V.S. clip-wise SED}\label{section:segment_clip_wise}
Section \ref{section:cnn} and \ref{section:cnn_transformer} introduce CNN, CNN-biGRU and CNN-Transformer architectures. In this section, we introduce how we apply the aforementioned architectures for AT and SED training with weakly labelled data. Conventional SED methods utilise strongly labelled data for supervised learning. However, collecting strongly labelled data is time consuming. The amount of strongly labelled data is therefore limited. To solve this problem, we propose to use weakly labelled dataset for SED. The SED systems with weakly labelled data can be categorized into segment-wise training \cite{Lee2017a, hershey2017cnn}, and our previously proposed clip-wise training \cite{xu2017large} methods. This section aims to investigate the comparison of the segment-wise and clip-wise training for AT and SED.

\subsection{Segment-wise training}
We denote the waveform of an audio clip as $ X $. For an $ X $ lasting for several seconds, we split it into several segments $ \{x_{m}\}_{m=1}^{M} $ where $ M $ is the number of segments. Each segment inherits the tags of the audio clip. We denote the tags of each segment as $ y \in \{0, 1\}^{K} $ where $ K $ is the number of sound classes. The SED problem is converted to an audio tagging problem on those segments. In training, a classifier $ f $ is trained on the segments. The loss function can be written as:
\begin{equation}
E = - \sum_{m=1}^{M} \sum_{k=1}^{K}[y_{k} \text{log}f(x_{m})_{k} + (1 - y_{k})\text{log}(1 - f(x_{m})_{k})].
\label{eq:segmentwise_loss}
\end{equation}

\noindent In inference, an audio clip is split into segments $ \{x_{m}\}_{m=1}^{M} $, and the SED result on each segment can be calculated by $ f(x_{m}) $. The AT result can be obtained by aggregating $ f(x_{m}) $ over all segments:
\begin{equation}
\label{eq:aggregation}
F(X) = \text{agg}(\{f(x_{m})\}_{m=1}^{M}). 
\end{equation}

\noindent The aggregation can be, for example, maximum or average operation over all the segments. The segment-wise classifier $ f $ can be CNN, CNN-biGRU or CNN-Transformer followed by a sigmoid non-linearity to predict the presence probability of sound events of each segment. We investigate the performance of choosing different duration of segments on SED and AT in Section VI.

\subsection{Clip-wise training}
In the segment-wise training, all segments $ x_{m} $ inherit the tags of an audio clip $ X $. The problem of segment-wise training is that many sound events may only last for a short time in the audio clip. Therefore, the tags of $ x_{m} $ may be incorrect because the segment may not contain the sound event. To solve this problem, our previous work proposed attention neural network based clip-wise training \cite{xu2017large}. The clip-wise training method does not explicitly assign tags for each segment $ x_{m} $. Instead, the systems are designed to learn the tags of $ x_{m} $ implicitly, that is, from the hidden layer of a neural network. We denote the segment-wise prediction of a segment $ x_{m} $ to be $ f(x_{m}) $. Then the prediction on the clip $ X $ can be obtained by aggregating the segment-wise predictions. For example, the aggregation can be a max, average or attention function over the prediction of all segments of each sound class. The max function can be defined as: 
\begin{equation}
F(X)_{k} = \underset{k}{\text{max}} f(x_{m})_{k}.
\label{eq:max}
\end{equation}
For example, the average function can be defined as:
\begin{equation}
F(X)_{k} = \sum_{m=1}^{M} f(x_{m})_{k}.
\label{eq:avg}
\end{equation}
The decision-level function can be defined as \cite{kong2019weakly}:
\begin{equation}
F(X)_{k} = \sum_{m=1}^{M} f(x_{m})_{k} p(x_{m})_{k},
\label{eq:att}
\end{equation}
where $ p(x_{m}) = \frac{\text{exp}(w(x_{m})_{k})}{\sum_{j=1}^{M}\text{exp}(w(x_{j})_{k})} $, and $ w(\cdot) $ is a linear transformation.
In training, we calculate the categorical binary crossentropy loss between the clip-level prediction $ F(X) $ and the ground truth label of $ X $:
\begin{equation}
E = - \sum_{k=1}^{K} \left [ y_{k} \text{log} F(X)_{k} + (1 - y_{k})\text{log}(1 - F(X)_{k}) \right ].
\label{eq:clipwise_loss}
\end{equation}
\noindent The difference between the clip-wise training (\ref{eq:clipwise_loss}) and the segment-wise training (\ref{eq:segmentwise_loss}) is that the clip-wise training directly outputs $ F(X) $, and can be trained in an end-to-end way with weakly labelled data. The $ f(x_{m})_{k} $ are latent representations learnt by the neural network.

\section{Automatic threshold optimization}\label{section:thresholding}
To obtain the presence or absence of sound events in an audio clip, AT systems need to apply thresholds to the system outputs. A sound class is predicted as presence if the AT output is larger than its corresponding threshold. The thresholds for AT are denoted as $ \Theta^{\text{AT}} = \{ \mu_{1}, ..., \mu_{K} \} $. Algorithm \ref{alg:at} shows the algorithm to obtain AT result from the AT system outputs.

\begin{algorithm}[t]
	\caption{Audio tagging}\label{alg:at}
	\begin{algorithmic}[1]
		\State Inputs: predicted presence probability of sound events in an audio clip $ F(X) $. AT thresholds $ \{ \mu_{1}, ..., \mu_{K} \} $.
	    \State Outputs: Predicted audio tags. 
	    \For {$k=1, ..., K$}
    	    \If {$ F(X)_{k} < \mu_{k} $} \State 
    	        return 0 for the $k$-th sound event.
    	    \Else \State
    	        return 1 for the $ k $-th sound event. 
    	    \EndIf
    	\EndFor
	\end{algorithmic}
\end{algorithm}

SED requires to predict not only the presence or absence but also the onset and offset times of sound events. \qk{Similar to AT, we first apply thresholds $ \{ \mu_{1}, ..., \mu_{K} \} $ on $ F(X) $ to predict the presence or absence of $ K $ classes of sound events in an audio clip $ X $. If the $ k $-th sound class is predicted to be present, then we apply a threshold $ \tau_{k}^{\text{high}} $ to the segment-wise predictions $ f(x_{m}) $ to detect sound events. In addition, to reduce the number of missed detection, a second threshold $ \tau_{k}^{\text{low}} $ is used. To begin with, we denote the neighbouring segments of an active segment as $ x' $. Then, a lower threshold $ \tau_{k}^{\text{low}} $ is applied on $ f(x_{m}) $ to obtain the calibrated sound event detection result. All thresholds for SED are denoted as $ \Theta^{\text{SED}} = \{ \mu_{1}, ..., \mu_{K}, \tau_{1}^{\text{high}}, ..., \tau_{K}^{\text{high}}, \tau_{1}^{\text{low}}, ..., \tau_{K}^{\text{low}} \} $. Algorithm \ref{alg:sed} summarizes obtaining the SED results from the clip-wise and segment-wise predictions.}

\begin{algorithm}[t]
	\caption{Sound event detection}\label{alg:sed}
	\begin{algorithmic}[1]
		\State Inputs: clip-wise prediction $ F(X) $, segment-wise prediction $ f(x_{m}) $, AT thresholds $ \{ \mu_{1}, ..., \mu_{K} \} $, SED high thresholds $ \{ \tau_{1}^{\text{high}}, ..., \tau_{K}^{\text{high}} \} $, SED low thresholds $ \{ \tau_{1}^{\text{low}}, ..., \tau_{K}^{\text{low}} \} $.
	    \State Outputs: Detected sound events. 
	    \For {$k=1, ..., K$}
    	    \If {$ F(X)_{k} < \mu_{k} $} \State return 0 for the $k$-th sound event.
    	    \Else 
    	        \For {$m = 1, ..., M$}
    	            \If {$ f(x_{m})_{k} > \tau_{k}^{\text{high}} $} \State
    	                Return 1 for the neighbouring segments $ x' $ of $ x_{m} $ if $ f(x')_{k} < \tau_{k}^{\text{low}} $. 
    	            \EndIf
    	        \EndFor
    	    \EndIf
    	\EndFor
	\end{algorithmic}
\end{algorithm}

The winning system of the AT subtask in DCASE 2017 Task 4 \cite{xu2017large} applies constant thresholds for both the AT and SED subtask. Setting those thresholds requires a lot of experience, and the manually selected thresholds are often not optimal. \qk{In addition, each sound class may have different thresholds. Therefore, sweeping over all combinations of thresholds is intractable.} We propose an automatic threshold optimization method to solve this problem. In the first stage, we optimize the systems and evaluate the systems based on the metrics that do not depend on the thresholds such as mean average precision (mAP). In the second stage, for a trained system, we optimize the thresholds over a specific metric such as F1 score or error rate (ER) to optimize the thresholds. 

For an audio clip $ X $, the AT result $ r^{\text{AT}} $ can be obtained by $ \text{alg}^{\text{AT}}(F(X), \Theta^{\text{AT}}) $ where $ \text{alg}^{\text{AT}} $ is the AT algorithm shown in Algorithm \ref{alg:at}. The SED result $ r^{\text{SED}} $ can be obtained by $ \text{alg}^{\text{SED}}(F(X), \{f(x_{m})\}_{m=1}^{M}, \Theta^{\text{SED}}) $ where $ \text{alg}^{\text{SED}} $ is the SED algorithm shown in Algorithm \ref{alg:sed}. The goal of AT or SED is to minimize some loss $ J(\Theta) $, for example, ER $ J_{\text{ER}}(\Theta) $ or negative F1 score $ J_{\text{F1}}(\Theta) $. The reason of using negative F1 is that minimizing $ J_{\text{F1}}(\Theta) $ is equivalent to maximizing F1 score. The optimization of thresholds becomes solving the following problem:

\begin{equation}
\hat{\Theta} = \underset{\Theta}{\text{argmin}} J(\Theta).
\label{eq:opt_threshold}
\end{equation}

\noindent The difficulty of solving (\ref{eq:opt_threshold}) is that $ \Theta $ consists of several parameters to be optimized. So applying grid search over all thresholds is inefficient. Another way is to use gradient based methods to iteratively optimize those thresholds. However, equation (\ref{eq:opt_threshold}) is a non-differentiable function, so we can not calculate the gradient over the thresholds in an analytical way. This is because both the AT and SED algorithms in Algorithm \ref{alg:at} and Algorithm \ref{alg:sed} contain non-differentiable operations such as thresholding. In addition, the evaluation metrics ER and F1 score are also non-differentiable. To solve this problem, we propose to calculate the gradients over the thresholds in a numerical way. That is, for each parameter $ \theta $, we calculate the gradient as:

\begin{equation}
\bigtriangledown_{\theta}J(\Theta) = \frac{J(\Theta + \bigtriangleup\Theta) - J(\Theta)}{\bigtriangleup\theta}, 
\label{eq:gradient}
\end{equation}

\noindent where $ \bigtriangleup\theta $ is a small constant number, and $ \bigtriangleup\Theta $ is a vector with all zero values the position of $ \theta $ which has a value of $ \bigtriangleup\theta $. After calculating the numerical gradient for all parameters $ \bigtriangledown_{\Theta}J = \{ \bigtriangledown_{\theta}J \}_{\theta \in \Theta} $, the optimized thresholds can be obtained by applying gradient based optimization methods iteratively: $ \Theta \leftarrow  \text{opt}(\Theta, \bigtriangledown_{\Theta}J) $, where opt denotes an optimization algorithm such as gradient descent (GD). GD optimization can be written by $ \Theta \leftarrow \Theta - \alpha \bigtriangledown_{\Theta}J $ where $ \alpha $ is a learning rate. \qk{We use Adam optimizer \cite{kingma2014adam} to optimize $ J(\Theta) $ due to its fast convergence. We describe Adam optimizer in Algorithm \ref{alg:adam} to show how it is used in our method. Overall, the automatic threshold optimization algorithm is described in Algorithm \ref{alg:opt_thresholds}. We have released our proposed automatic threshold optimization toolbox called \textit{autoth}\footnote{\url{https://github.com/qiuqiangkong/autoth}}}.

\begin{algorithm}[t]
	\caption{Adam optimization. Symbol $ g_{t}^{2} $ indicates the elementwise square $ g_{t} \odot g_{t} $. Learning rate is denoted as $ \alpha $. Hyper-parameters are set to $ \beta_{1}=0.9 $, $ \beta_{2}=0.999 $ and $ \epsilon=10^{-8} $ following \cite{kingma2014adam}. }\label{alg:adam}
	\begin{algorithmic}[1]
		\State Inputs: parameters $ \Theta $.
	    \State Init $ \Theta_{0}, m_{0}=0, v_{0}=0, t=0 $
		\While {$\Theta$ not converged}
    		\State $ t \leftarrow t + 1 $
    	    \State $ g_{t} = \bigtriangledown_{\Theta}J $
    	    \State $ m_{t} \leftarrow \beta_{1}m_{t-1} + (1 - \beta_{1})g_{t} $
    	    \State $ v_{t} \leftarrow \beta_{2}v_{t-1} + (1 - \beta_{2})g_{t}^2 $
    	    \State $\hat{m}_{t} \leftarrow m_{t} / (1 - \beta_{1}^{t})$
    	    \State $\hat{v}_{t} \leftarrow v_{t} / (1 - \beta_{2}^{t})$
    	    \State $ \Theta_{t} \leftarrow \Theta_{t-1} - \alpha \cdot \hat{m}_{t} / (\sqrt{\hat{v}_{t}} + \epsilon) $
        \EndWhile
	\end{algorithmic}
\end{algorithm}

\begin{algorithm}[t]
	\caption{Automatic thresholds optimization}\label{alg:opt_thresholds}
	\begin{algorithmic}[1]
		\State Inputs: Validation dataset $ D = \{X^{(n)}, y^{(n)}\}_{n=1}^{N} $, trained AT system $ F(\cdot) $, trained SED system $ f(\cdot) $.
	    \State Outputs: Optimized thresholds $ \Theta $.
	    \State Initialize $ \Theta $.
	    \For {$ i=1, ..., \text{ITER} $}
    	    \For {$ n=1, ..., N $} \State
    	        $ \hat{y}^{(n)} = \text{alg}(F(X^{(n)}), f(x_{m}^{(n)}), \Theta) $. 
    	    \EndFor
    	    \State $ J = \text{metric}(\{ \hat{y}^{(n)} \}_{n=1}^{n=N}, \{ y^{(n)} \}_{n=1}^{n=N}) $ 
    	    
    	    \For {$ \theta $ in $ \Theta $} \State
        	    $\bigtriangledown_{\theta}J = \frac{J(\Theta + \bigtriangleup\Theta) - J(\theta)}{\bigtriangleup\theta}$
        	\EndFor
        	\State $ \bigtriangledown_{\Theta}J = \{ \bigtriangledown_{\theta}J \}_{\theta \in \Theta} $
        	\State $ \Theta \leftarrow  \text{opt}(\Theta, \bigtriangledown_{\Theta}J) $
        \EndFor
	\end{algorithmic}
\end{algorithm}

\section{Experiments}

\subsection{Experimental setup}\label{section:experiments}
There are several SED datasets including the DCASE 2017 Task 4 \cite{dcase2017task4}, the DCASE 2018 Task 4 \cite{serizel2018large} and the DCASE 2019 Task 4 \cite{turpault2019sound}. 
We evaluate our SED system on the DCASE 2017 Task 4 ``large-scale weakly supervised sound event detection for smart cars''. The reason we choose this dataset is because it is a large-scale dataset containing over 140 hours of weakly labelled audio clips for training. The audio recordings of the DCASE 2017 Task 4 are from a subset of AudioSet \cite{audioset} where each audio clip is extracted from YouTube video. DCASE 2017 Task 4 consists of 17 sound events divided into two categories: ``warning'' and ``vehicle''. Most of those audio clips have duration of 10 seconds. The audio clips shorter than 10 seconds are padded with silence to 10 seconds. Table \ref{tab:event_name} lists the sound events and their statistics. The DCASE 2017 Task 4 dataset consists of a training subset with 51172 audio clips, a validation subset with 488 audio clips, and an evaluation set with 1103 audio clips. The training subset is weakly labelled. The validation and evaluation subsets are both weakly and strongly labelled for evaluation. The source code of this work is released\footnote{\url{https://github.com/qiuqiangkong/sound_event_detection_dcase2017_task4}}.

\begin{table}[t]  
	\centering
	\caption{Sound events in the DCASE 2017 Task 4 ``large-scale weakly supervised sound event detection for smart cars''}
\begin{tabular}{c|lr}
	\hline
	& \textbf{Event name} & \multicolumn{1}{l}{\textbf{Training number}} \\
	\hline
	\hline
	\multirow{9}[0]{*}{\textcolor[rgb]{ .133,  .133,  .133}{\textbf{Warning sounds}}} & \textcolor[rgb]{ .133,  .133,  .133}{Train horn} & 441 \\
	& \textcolor[rgb]{ .133,  .133,  .133}{Air horn, truck horn} & 407 \\
	& \textcolor[rgb]{ .133,  .133,  .133}{Car alarm} & 273 \\
	& \textcolor[rgb]{ .133,  .133,  .133}{Reversing beeps} & 337 \\
	& \textcolor[rgb]{ .133,  .133,  .133}{Ambulance (siren)} & 624 \\
	& \textcolor[rgb]{ .133,  .133,  .133}{Police car (siren)} & 2399 \\
	& \textcolor[rgb]{ .133,  .133,  .133}{Fire engine, fire truck (siren)} & 2399 \\
	& \textcolor[rgb]{ .133,  .133,  .133}{Civil defense siren} & 1506 \\
	& \textcolor[rgb]{ .133,  .133,  .133}{Screaming} & 744 \\
	\hline
	\multirow{8}[0]{*}{\textcolor[rgb]{ .133,  .133,  .133}{\textbf{Vehicle sounds}}} & \textcolor[rgb]{ .133,  .133,  .133}{Bicycle} & 2020 \\
	& \textcolor[rgb]{ .133,  .133,  .133}{Skateboard} & 1617 \\
	& \textcolor[rgb]{ .133,  .133,  .133}{Car} & 25744 \\
	& \textcolor[rgb]{ .133,  .133,  .133}{Car passing by} & 3724 \\
	& \textcolor[rgb]{ .133,  .133,  .133}{Bus} & 3745 \\
	& \textcolor[rgb]{ .133,  .133,  .133}{Truck} & 7090 \\
	& \textcolor[rgb]{ .133,  .133,  .133}{Motorcycle} & 3291 \\
	& \textcolor[rgb]{ .133,  .133,  .133}{Train} & 2301 \\
	\hline
\end{tabular}%
	\label{tab:event_name}
\end{table}

\subsection{Feature}
We use log mel spectrogram as input feature following previous work on audio tagging \cite{choi2016automatic, hershey2017cnn, kong2019cross}. To begin with, all audio clips are converted to monophonic and resampled to 32 kHz. The short time Fourier transform with a Hanning window of 1024 samples and a hop size of 320 samples is used to extract spectrogram which leads to 100 frames in a second. We apply 64 mel filter banks on the spectrogram followed by logarithmic operation to calculate log mel spectrogram. The number 64 is chosen so that it can be evenly divided by a power of 2 in the down-sampling layers of CNNs. The mel filter banks have a lower cut-off frequency of 50 Hz and a higher cut-off frequency of 14 kHz to avoid aliasing caused by resampling.

\subsection{Model}
The segment-wise training systems are described in equation (\ref{eq:aggregation}), and are modeled by a 9-layer CNN which has shown to perform well on a variety of audio tagging tasks \cite{kong2019cross}. Table \ref{tb:architecture} shows that the 9-layer CNN consists of 4 convolutional blocks, where each convolutional block consists of 2 convolutional layers with kernel sizes of $ 3 \times 3 $. Batch normalization \cite{ioffe2015batch} and ReLU non-linearity \cite{dahl2013improving} is applied after each convolutional layer. The convolutional block consists of 64, 128, 256 and 512 feature maps, respectively. A $ 2 \times 2 $ average pooling is applied after each convolutional block to extract high level features. \qk{We did not apply residual connections in our CNNs as gradient vanishing is not a problem with 8 convolutional layers.} In Table \ref{tb:architecture}, the number following $ @ $ represents the number of feature maps. The second column shows the number of batch size (bs), feature maps, frames and frequency bins. We average out the frequency axis of the output from the last convolutional layer. Then time distributed fully connected layer with sigmoid non-linearity is applied to predict the presence probability of sound events of each time frame. To obtain the AT result for supervised learning, aggregation functions including max, average and attention along time frames are applied. Adam \cite{kingma2014adam} optimizer with a learning rate of 0.001 is applied, and is reduced to 0.0001 after the performance is plateaued on validation data. Mixup \cite{zhang2017mixup} with alpha of 1.0 is used in all experiments to prevent training from overfitting. \qk{The training is stopped at 60,000 iterations by observing the performance on the validation set.}

\begin{table}[]
\caption{CNN architecture.}
\label{tb:architecture}

\begin{tabular}{c|c|c}
\hline
Layers & Output size & Param. num. \\ \hline
Input: log mel spectrogram & $ \text{bs} \times 1 \times 640 \times 64 $ & - \\ \hline
$ \begin{pmatrix} 3 \times 3 \ @ \ 64 \\ \text{BN, ReLU} \end{pmatrix} \times 2 $ & $ \text{bs} \times 64 \times 640 \times 64 $ & 37,696 \\ \hline
$ 2 \times 2 $ avg. pooling & $ \text{bs} \times 64 \times 320 \times 32 $ & - \\ \hline
$ \begin{pmatrix} 3 \times 3 \ @ \ 128 \\ \text{BN, ReLU} \end{pmatrix} \times 2 $ & $ \text{bs} \times 128 \times 320 \times 32 $ & 221,696 \\ \hline
$ 2 \times 2 $ avg. pooling & $ \text{bs} \times 128 \times 160 \times 16 $ & - \\ \hline
$ \begin{pmatrix} 3 \times 3 \ @ \ 256 \\ \text{BN, ReLU} \end{pmatrix} \times 2 $ & $ \text{bs} \times 256 \times 160 \times 16 $ & 885,760 \\ \hline
$ 2 \times 2 $ avg. pooling & $ \text{bs} \times 256 \times 80 \times 8 $ & - \\ \hline
$ \begin{pmatrix} 3 \times 3 \ @ \ 512 \\ \text{BN, ReLU} \end{pmatrix} \times 2 $ & $ \text{bs} \times 512 \times 80 \times 8 $ & 3,540,992 \\ \hline
Embedding: Avg. out freq. bins & $ \text{bs} \times 512 \times 80 \times 1 $ & - \\ \hline
\end{tabular}
\end{table}

\subsection{Evaluation metrics}
To evaluate the systems performance, we use the precision, recall and F1 score which are described in \cite{mesaros2016metrics}:
\begin{equation}
\text{P}=\frac{\text{TP}}{\text{TP}+\text{FP}}
\end{equation}
\begin{equation}
\text{R}=\frac{\text{TP}}{\text{TP}+\text{FN}}
\end{equation}
\begin{equation}
\text{F1}=\frac{2 \text{P} \cdot \text{R}}{\text{P} + \text{R}},
\end{equation}
where TP, FP, FN are the number of true positive, false positive and false negative samples, respectively. The higher precision, recall and F1 score indicate better performance. Usually thresholds need to be manually selected and applied on the system outputs to calculate TP, FP and FN. We use average precision (AP) metric \cite{audioset} to compare the performance of different systems because the AP does not depend on thresholds. AP is defined as the area under the precision-recall curve calculated at multiple thresholds. Mean average precision (mAP) is the averaged AP over all sound classes. The higher mAP indicates better performance. Random guess has an mAP of 0.06 for the DCASE 2017 Task 4 containing 17 sound classes. The mAP is a \textit{macro-averaging} statistic because it is calculated independently within a sound class. Then, the statistics are averaged across all sound classes. On the other hand, \textit{micro-averaging} statistic is calculated from outputs and ground truths flattened from all classes. 

For the AT subtask, systems are ranked based on macro-averaging F1 score. For the SED subtask, systems are ranked based on micro-averaging F1 score and Error rate (ER) evaluated on 1-second segments \cite{mesaros2017dcase}. \qk{Error rate measures the amount of errors in terms of insertions (I), deletions (D) and substitutions (S), and is defined as follows \cite{mesaros2016metrics}:
\begin{equation}
\text{ER}=\frac{\sum_{m} S(m)+\sum_{m} D(m) + \sum_{m} I(m)}{\sum_{m} N(m)}, 
\end{equation}
where $ I(m), D(m), S(m), N(m) $ are the number of inserted, deleted, substituted, and ground truth sound events in the $ m $-th segment. Lower ER indicates better performance.} The segment based evaluation is calculated in a fixed time grid, using segments of one second length to compare the ground truth and the system output \cite{dcase2017task4}. Similarly, segment based F1-score are calculated in the same way.

%More sound event detection demos can also be found on the web\footnote{\url{https://sites.google.com/view/xuyong/demos/dcase2017_task4_sed}}.

\begin{table*}[t]
\centering
\caption{The percentage of time containing sound events in an audio clip labelled as containing the sound event.}
\label{table:duration_portion}
\begin{tabular}{*{18}{c}}
 \toprule
 Train horn & Air horn, truck horn & Car alarm & Reversing beeps & Bicycle & Skateboard & Ambulance & Fire engine, fire truck & Civil defense siren \\
 \midrule
 0.400 & 0.553 & 0.463 & 0.538 & 0.478 & 0.616 & 0.829 & 0.905 & 0.930 \\
 \bottomrule
 \toprule
   Police car & Screaming & Car & Car passing by & Bus & Truck & Motorcycle & Train & Avg \\
 \midrule
 0.746 & 0.619 & 0.606 & 0.768 & 0.885 & 0.764 & 0.696 & 0.776 & 0.681 \\
 \bottomrule
\end{tabular}
\end{table*}

\begin{figure*}[t]
	\centering
	\centerline{\includegraphics[width=\textwidth]{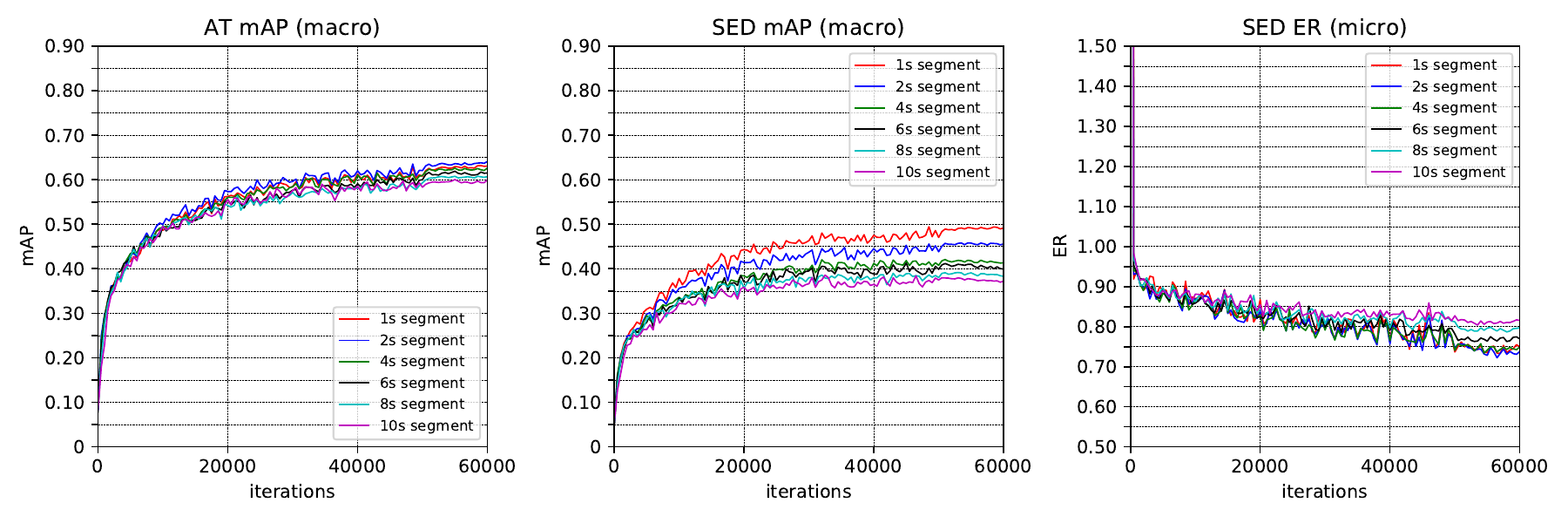}}
	\caption{Segment-wise training result trained with different durations of audio segments. From left to right: Audio tagging macro mAP; SED macro mAP; SED micro error rate. }
	\label{fig:segment_metrics}
\end{figure*}

\subsection{Segment-wise AT and SED}
\qk{There is a lack of research comparing segment-wise \cite{hershey2017cnn} and clip-wise \cite{xu2017large} training for AT and SED. We first investigate the segment-wise training method. To begin with, an audio clip is split into segments. Then SED predictions are calculated by running audio tagging system on segments.} Because the audio clips are weakly labelled, there is no information when a sound event occurs and how long they last. This can affect the label accuracy of segment-wise training because all segments inherit the tags from the audio clip. \qk{In inference, the SED result is obtained by predicting audio tags on segments.} \qk{Table \ref{table:duration_portion} shows the average percentage of time frames in an audio clip containing different sound events from the validation set of DCASE 2017 Task 4.} Sound events such as civil defense siren has a presence percentage of 0.930 which indicates segment-wise labels are more likely to be correct. Sound events such as train horn has a presence percentage of 0.400 which indicates segment-wise label is less likely to be correct.

The 10-second audio clips are split into segments with different lengths from 1 to 10 seconds. Each segment inherit the tags from the audio clip. \qk{The minimum 1-second setting follows \cite{hershey2017cnn}.} A 9-layer CNN is applied to build the segment-wise training systems. Fig. \ref{fig:segment_metrics} shows the mAP and ER of AT and SED with different segment durations. Training with 2-second segments achieves an mAP of 0.64 in audio tagging, slightly outperforming other segment duration in AT. This indicates that the prediction of long segments does not perform well when no attention or temporal dependency is used. The second column of Fig. \ref{fig:segment_metrics} shows that training with 1-second segments achieves an SED mAP of 0.44, outperforming other segment duration. \qk{This indicates that shorter segments achieve better SED result than longer segments when using segment-wise training systems. One explanation is that SED is obtained by AT on segments, so AT on shorter segments can provide higher SED resolution.} To calculate the ER, we use constant AT thresholds of $ \mu_{k}=0.5, k=1,...,K $, and SED thresholds of $ \tau_{k}^{\text{high}}=0.3, \tau_{k}^{\text{low}}=0.1, k=1,...,K $ for all sound class following \cite{xu2017large}. The third column of Fig. \ref{fig:segment_metrics} shows that the 1-second and 2-second segment duration achieve an ER of 0.74, outperforming other segment durations. 

\begin{figure*}[t]
	\centering
	\centerline{\includegraphics[width=\textwidth]{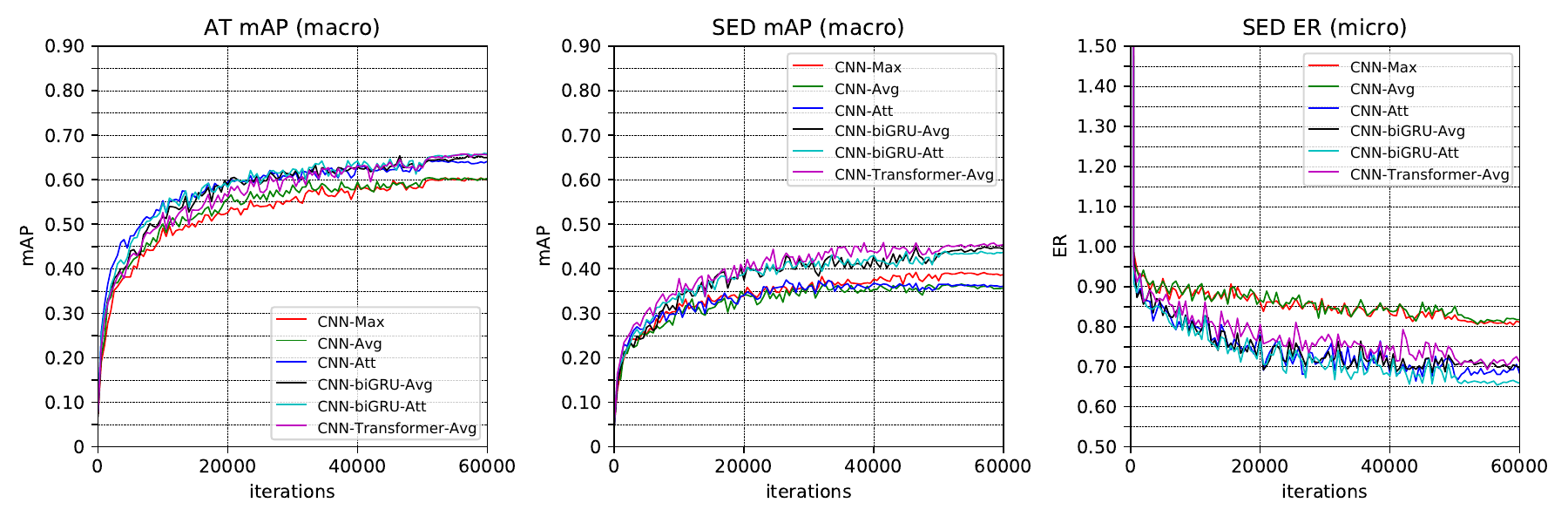}}
	\caption{Clip-wise training result with different systems. Audio tagging macro mAP; SED macro mAP; SED micro error rate. }
	\label{fig:clipwise_at_sed_test}
\end{figure*}

\begin{figure*}[t]
	\centering
	\centerline{\includegraphics[width=\textwidth]{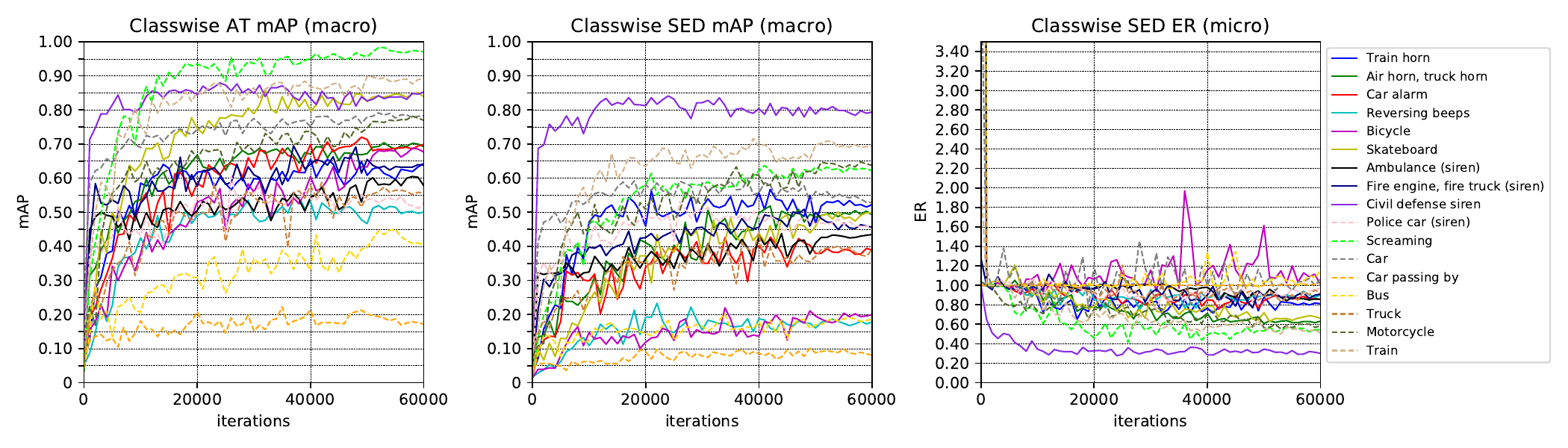}}
	\caption{From left to right: Class-wise AT average precision; Class-wise SED average precision; Class-wise SED error rate with the CNN-biGRU-Att system.}
	\label{fig:best_model_17_classes_evaluate}
\end{figure*}

\begin{figure*}[t]
	\centering
	\centerline{\includegraphics[width=18cm]{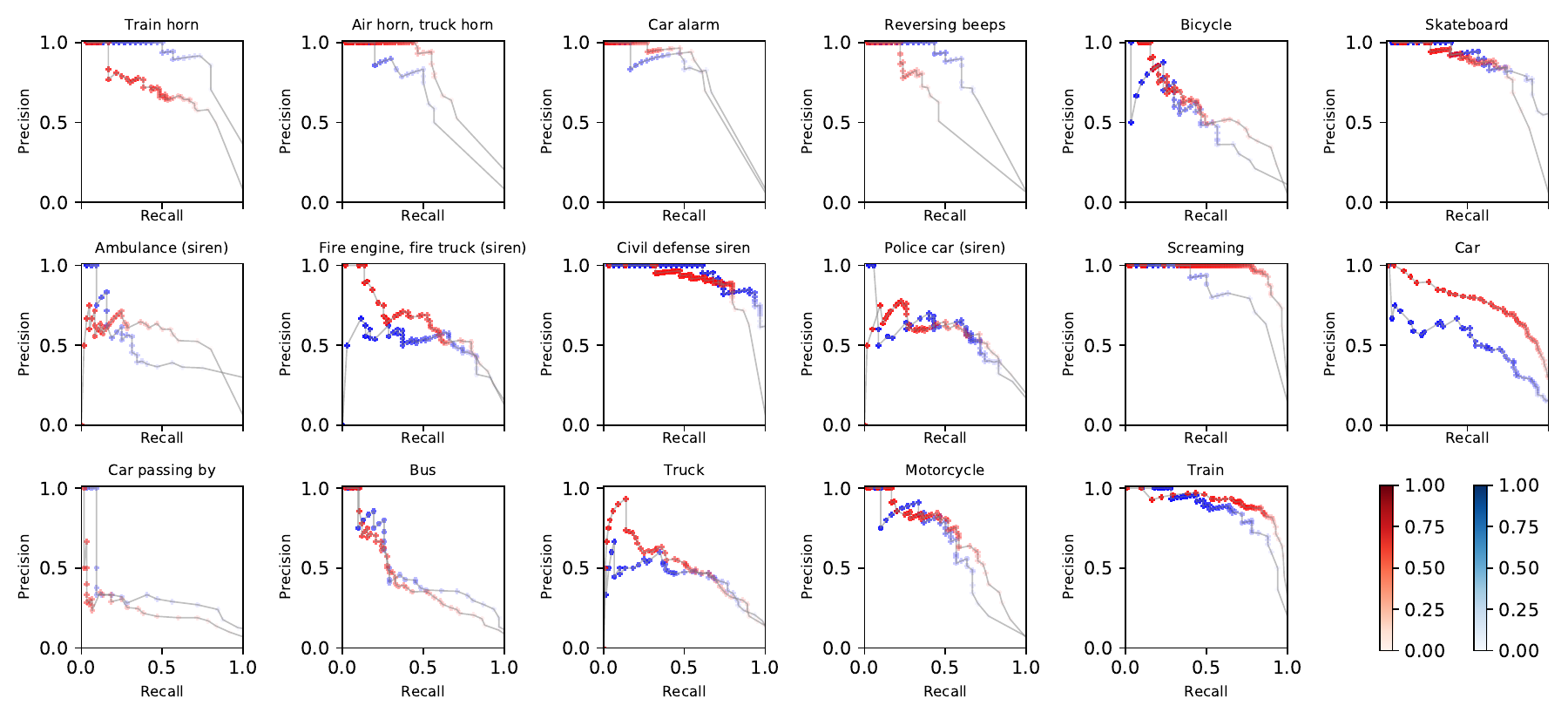}}
	\caption{Precision-recall curves of sound classes at different thresholds with the CNN-biGRU-Att system.}
	\label{fig:at_thresholds}
\end{figure*}

\begin{table*}
  \caption{Performance of the proposed systems on the validation (Val.) and evaluation (Eval.) subset}
  \vspace{6pt}
  \label{tab:result}
  \centering
  \begin{tabular}{l p{.6cm}p{.6cm}p{.6cm}p{.6cm}p{.6cm}p{.6cm}p{.6cm}|p{.6cm}p{.6cm}p{.6cm}p{.6cm}p{.6cm}p{.6cm}p{.6cm}}
    \toprule
    & \multicolumn{4}{c}{\textbf{\textsc{Val. AT}}} & \multicolumn{3}{c|}{\textbf{\textsc{Val. SED}}} & \multicolumn{4}{c}{\textbf{\textsc{Eval. AT}}} & \multicolumn{3}{c}{\textbf{\textsc{Eval. SED}}} \\
    \cmidrule(lr){2-5} \cmidrule(lr){6-8} \cmidrule(lr){9-12} \cmidrule(lr){13-15}
    Systems & mAP & F1 & P & R & mAP & F1 & ER & mAP & F1 & P & R & mAP & F1 & ER \\
    \midrule
    
    CNN-Max & 0.579 & 0.479 & 0.497 & 0.462 & 0.359 & 0.352 & 0.83 & 0.602 & 0.565 & 0.593 & 0.539 & 0.387 & 0.466 & 0.75 \\
    CNN-Max (auto thres) & \# & 0.538 & 0.633 & 0.467 & \# & 0.489 & 0.74 & \# & 0.582 & 0.674 & 0.513 & \# & 0.520 & 0.72 \\
    CNN-Avg & 0.587 & 0.479 & 0.495 & 0.464 & 0.342 & 0.346 & 0.83 & 0.603 & 0.553 & 0.578 & 0.530 & 0.357 & 0.452 & 0.76 \\
    CNN-Avg (auto thres) & \# & 0.564 & \textbf{0.649} & 0.498 & \# & 0.357 & 0.76 & \# & 0.597 & 0.665 & 0.542 & \# & 0.518 & 0.74 \\
    CNN-Att & 0.637 & 0.559 & 0.557 & 0.561 & 0.339 & 0.480 & 0.82 & 0.641 & 0.605 & 0.607 & 0.604 & 0.361 & 0.552 & 0.80 \\
    CNN-Att (auto thres) & \# & \textbf{0.604} & 0.596 & \textbf{0.612} & \# & 0.523 & 0.68 & \# & 0.617 & 0.610 & 0.624 & \# & 0.545 & 0.69 \\
    CNN-biGRU-Avg & 0.650 & 0.555 & 0.541 & 0.569 & \textbf{0.456} & 0.491 & 0.90 & 0.650 & 0.632 & 0.638 & 0.627 & 0.444 & 0.564 & 0.84 \\
    CNN-biGRU-Avg (auto thres) & \# & 0.602 & 0.609 & 0.594 & \# & 0.534 & 0.685 & \# & 0.632 & 0.648 & 0.617 & \# & 0.567 & 0.72 \\
    CNN-biGRU-Att & 0.647 & 0.557 & 0.554 & 0.559 & 0.419 & 0.492 & 0.80 & 0.658 & 0.625 & 0.632 & 0.617 & 0.436 & 0.564 & 0.78 \\
    CNN-biGRU-Att (auto thres) & \# & 0.581 & 0.575 & 0.587 & \# & \textbf{0.537} & \textbf{0.65} & \# & 0.640 & 0.637 & \textbf{0.642} & \# & \textbf{0.584} & \textbf{0.68} \\
    CNN-Transformer-Avg & \textbf{0.653} & 0.557 & 0.554 & 0.561 & 0.437 & 0.483 & 0.91 & \textbf{0.656} & 0.629 & 0.644 & 0.616 & \textbf{0.454} & 0.556 & 0.87 \\
    CNN-Transformer-Avg (auto thres) & \# & 0.599 & 0.636 & 0.566 & \# & 0.524 & 0.75 & \# & \textbf{0.646} & \textbf{0.691} & 0.607 & \# & 0.573 & 0.75 \\
    \bottomrule
\end{tabular}
\end{table*}

\begin{table}[h]  
	\centering
	\caption{AT of different systems}
\begin{tabular}{lccc}
	\hline
	\textbf{Dev-set} & \textbf{F1} & \textbf{Precision} & \textbf{Recall} \\
	\hline
	DCASE2017 Baseline \cite{mesaros2017dcase} & 0.109  & 0.078   & 0.175 \\
	MIL-NN ensemble \cite{tseng2017multiple} & 0.353  & 0.286  & 0.460 \\
	CNN-ensemble \cite{Lee2017a} & 0.570 & \textbf{0.703} & 0.479 \\
	Previously submitted fusion \cite{xu2017large} & 0.577  & 0.565  & 0.589 \\
	CNN-biGRU-Att (auto thres) & 0.581 & 0.575 & 0.587 \\
	CNN-Transformer-Avg (auto thres) & \textbf{0.599} & 0.636 & 0.566 \\
	\hline
	\hline
	\textbf{Eval-set} & \textbf{F1} & \textbf{Precision} & \textbf{Recall} \\
	\hline
	DCASE2017 Baseline \cite{mesaros2017dcase} & 0.182  & 0.150  & 0.231 \\
	MIL-NN \cite{tseng2017multiple} & 0.352  & 0.316  & 0.397 \\
	CNN-ensemble \cite{Lee2017a} & 0.526  & \textbf{0.697} & 0.423 \\
	Previously submitted fusion system \cite{xu2017large} & 0.556  & 0.614  & 0.508 \\
	CNN-biGRU-Att (auto thres) & 0.640 & 0.637 & \textbf{0.642} \\
	CNN-Transformer-Avg (auto thres) & \textbf{0.646} & 0.691 & 0.607 \\
	\hline
\end{tabular}%
	\label{tab:at}
\end{table}

\begin{table}[h]  
	\centering
	\caption{SED results of different systems}
% Table generated by Excel2LaTeX from sheet 'Sheet1'
\begin{tabular}{lcc}
	\hline
	\textbf{Dev-set} & \textbf{F1} & \textbf{Error rate} \\
	\hline
	DCASE2017 baseline \cite{mesaros2017dcase} & 0.138  & 1.02 \\
	CNN-ensemble \cite{Lee2017a} & 0.471  & \textbf{0.71} \\
	MultiScale-CNN \cite{Lee2017} & 0.342  & 0.84 \\
	Previously submitted fusion \cite{xu2017large} & \textbf{0.497}  & 0.72 \\
	CNN-biGRU-Att (auto thres) & \textbf{0.537} & \textbf{0.65} \\
	CNN-Transformer-Avg (auto thres) & 0.524 & 0.75 \\
	\hline
	\hline
	\textbf{Eval-set} & \textbf{F1} & \textbf{Error rate} \\
	\hline
	DCASE2017 baseline \cite{mesaros2017dcase} & 0.284  & 0.93 \\
	CNN-ensemble \cite{Lee2017a} & 0.555  & \textbf{0.66} \\
	MultiScale-CNN \cite{Lee2017} & 0.471  & 0.75 \\
	Previously submitted fusion \cite{xu2017large} & 0.518  & 0.73 \\
	CNN-biGRU-Att (auto thres) & \textbf{0.584} & 0.68 \\
	CNN-Transformer-Avg (auto thres) & 0.573 & 0.75 \\
	\hline
\end{tabular}%
	\label{tab:sed}
\end{table}

\subsection{Clip-wise AT and SED}
We investigate the clip-wise training systems in this section. The difference of the clip-wise training and the segment-wise training is that with clip-wise training, the SED result can be obtained from the intermediate layer of a neural network. Then, the AT predictions can be calculated by the aggregation functions such as (\ref{eq:max}, \ref{eq:avg}, \ref{eq:att}). FIG. \ref{fig:clipwise_at_sed_test} shows the AT and SED performance of the clip-wise CNN systems. For the AT subtask, \qk{the decision-level maximum (CNN-Max) or decision-level average (CNN-Avg) systems achieve an mAP of 0.60. The decision-level attention (CNN-Att) improves the mAP to 0.64, indicating the attention plays an important role in AT. The CNN-biGRU systems and the CNN-Transformer system further improve the mAP performance to 0.65}, indicating that the temporal dependency information is helpful for AT. For the SED subtask, \qk{the CNN-Transformer system achieves an SED mAP of 0.45, slightly outperforming the CNN-biGRU systems of 0.44 and other CNN systems of 0.36 to 0.39, respectively. On the other hand, CNN-biGRU achieves an ER of 0.66, outperforming other systems ranging from 0.69 to 0.86}. To calculate ER we applies thresholds that are the same as the segment-wise training systems. \qk{Fig. \ref{fig:segment_metrics} and Fig. \ref{fig:clipwise_at_sed_test} show that the segment-wise training achieves better mAP on the SED task, while the clip-wise training achieves better ER on the SED task. One explanation is that mAP is evaluated in frame-wise, while ER is evaluated in 1-second segments.}

Fig. \ref{fig:best_model_17_classes_evaluate} shows the class-wise performance of the CNN-biGRU-Att system over training iterations. The performance on different sound classes varies. The prediction of screaming achieves the highest AT mAP of 0.94. On the other hand, the prediction of car passing by achieves the lowest mAP of 0.20. One explanation of the underperformance of sound classes such as car passing by is that they are difficult to perceive even by human in audio recordings. For SED, some sound classes achieve better mAP than others, for example, civil defense siren achieves the highest mAP of 0.80, indicating the system is performing well on these sound classes. The ER curve of different sound classes is different. Civil defense siren has an ER of 0.26 while other sound classes such as bicycle has ER over 1. The class-wise results show that both the AT and SED performance vary from sound class to sound class.

\subsection{Automatic thresholds optimization}
Previous subsection shows that the performance of different sound classes can be different. It can be useful to observe their precision-recall curves under various thresholds. \qk{Fig. \ref{fig:at_thresholds} shows the AT precision-recall curve of sound classes with the CNN-biGRU-Att system under different thresholds ranging from 0 and 1.} The blue and red curve show the validation and evaluation precision-recall curve, respectively. Fig. \ref{fig:at_thresholds} shows that the validation and the evaluation curve have similar trend but are not overlapped, indicating that the data distribution of validation and evaluation data can be slightly different. Some sound classes such as screaming have high precision at a variety of thresholds. On the other hand, the precision drops rapidly with the increase of recall for some sound classes such as car passing by. Fig. \ref{fig:at_thresholds} shows that different sound classes have different thresholds to achieve optimal metrics such as F1 score.

Table \ref{tab:result} shows the AT and SED performance with the clip-wise training systems. We first apply constant thresholds for both AT and SED systems. The constant thresholds are the same as the thresholds applied in previous sections. In addition to applying the constant thresholds, we apply thresholds $ \Theta^{\text{AT}} $ and $ \Theta^{\text{SED}} $ for AT and SED obtained by using the automatic thresholds optimization algorithm described in Algorithm \ref{alg:opt_thresholds}. Table \ref{tab:result} shows that the proposed automatic thresholds optimization improved both the AT and SED performance. For example, the CNN-Transformer AT F1 score improves from 0.557 to 0.599, and 0.629 to 0.646 in the validation and evaluation set, respectively. The CNN-biGRU SED ER is reduced from 0.80 to 0.65, and 0.78 to 0.68 in the validation and evaluation set, respectively. Those results show the effectiveness of the proposed automatic threshold optimization method.

Table \ref{tab:at} shows the precision, recall and F1-score of different methods for the AT on the validation and evaluation sets, respectively. The official DCASE2017 baseline is give in \cite{mesaros2017dcase} by using a multilayer perceptron (MLP) classifier, denoted as ``DCASE2017 Baseline''. The MIL-NN is a multiple instance learning based neural network system proposed in \cite{tseng2017multiple}. The CNN-ensemble system is proposed by \cite{Lee2017a} and ranked the 1st in the SED subtask in Task 4 of the DCASE 2017 challenge. Our proposed systems achieve an F1 score of 0.646 on the evaluation set, outperforming the other methods in AT. The CNN-biGRU and the CNN-Transformer systems achieve similar performance. Table \ref{tab:sed} shows the SED result with different methods. On the evaluation set, our proposed CNN-biGRU-Att with automatic thresholds optimization achieves an F1 score of 0.584, outperforming other systems. The system achieves an ER of 0.68 which is comparable with the ensemble based CNN system \cite{Lee2017a}.

\section{Conclusion} \label{section:conclusion}
This paper investigates sound event detection (SED) with weakly labelled data. The variants of convolutional neural networks (CNNs) and CNN-Transformer systems were proposed for the audio tagging and sound event detection. The segment-wise training and clip-wise training systems were investigated. The clip-wise training achieves an mAP of 0.650 in audio tagging and an ER of 0.68 in SED. A novel automatic thresholds optimization method is proposed to approach the thresholds selection problem. The automatic thresholds optimization method improves the AT F1 score from 0.629 to 0.646, and reduces the ER from 0.78 to 0.68. \qk{We show that the CNN-Transformer performs similarly to the CRNN system, while the CNN-Transformer has the advantage of being computed in parallel. In addition, the improvements of audio tagging and SED are mainly from the automatic threshold optimization.} In future, we will focus on extending the sound event detection systems to large-scale training data such as AudioSet.

\section{Acknowledgement}
Qiuqiang Kong was supported by EPSRC grant EP/N014111/1 ``Making Sense of Sounds'' and a Research Scholarship from the China Scholarship Council (CSC) No. 201406150082.

% if have a single appendix:
%\appendix[Proof of the Zonklar Equations]
% or
%\appendix  % for no appendix heading
% do not use \section anymore after \appendix, only \section*
% is possibly needed

% use appendices with more than one appendix
% then use \section to start each appendix
% you must declare a \section before using any
% \subsection or using \label (\appendices by itself
% starts a section numbered zero.)
%

%\appendices
%\section{Proof of the First Zonklar Equation}
%Appendix one text goes here.

% you can choose not to have a title for an appendix
% if you want by leaving the argument blank
%\section{}
%Appendix two text goes here.

% Can use something like this to put references on a page
% by themselves when using endfloat and the captionsoff option.
\ifCLASSOPTIONcaptionsoff
  \newpage
\fi

% trigger a \newpage just before the given reference
% number - used to balance the columns on the last page
% adjust value as needed - may need to be readjusted if
% the document is modified later
%\IEEEtriggeratref{8}
% The "triggered" command can be changed if desired:
%\IEEEtriggercmd{\enlargethispage{-5in}}

% references section

% can use a bibliography generated by BibTeX as a .bbl file
% BibTeX documentation can be easily obtained at:
% http://mirror.ctan.org/biblio/bibtex/contrib/doc/
% The IEEEtran BibTeX style support page is at:
% http://www.michaelshell.org/tex/ieeetran/bibtex/
%\bibliographystyle{IEEEtran}
% argument is your BibTeX string definitions and bibliography database(s)
%\bibliography{IEEEabrv,../bib/paper}
%
% <OR> manually copy in the resultant .bbl file
% set second argument of \begin to the number of references
% (used to reserve space for the reference number labels box)
%\begin{thebibliography}{1}
%\bibitem{IEEEhowto:kopka}
%H.~Kopka and P.~W. Daly, \emph{A Guide to \LaTeX}, 3rd~ed.\hskip 1em plus
%  0.5em minus 0.4em\relax Harlow, England: Addison-Wesley, 1999.
%\end{thebibliography}

\bibliographystyle{IEEEtran}
\bibliography{refs}

% biography section
% 
% If you have an EPS/PDF photo (graphicx package needed) extra braces are
% needed around the contents of the optional argument to biography to prevent
% the LaTeX parser from getting confused when it sees the complicated
% \includegraphics command within an optional argument. (You could create
% your own custom macro containing the \includegraphics command to make things
% simpler here.)
%\begin{IEEEbiography}[{\includegraphics[width=1in,height=1.25in,clip,keepaspectratio]{mshell}}]{Michael Shell}
% or if you just want to reserve a space for a photo:

\begin{IEEEbiography}[{\includegraphics[width=1in,height=1.25in,clip,keepaspectratio]{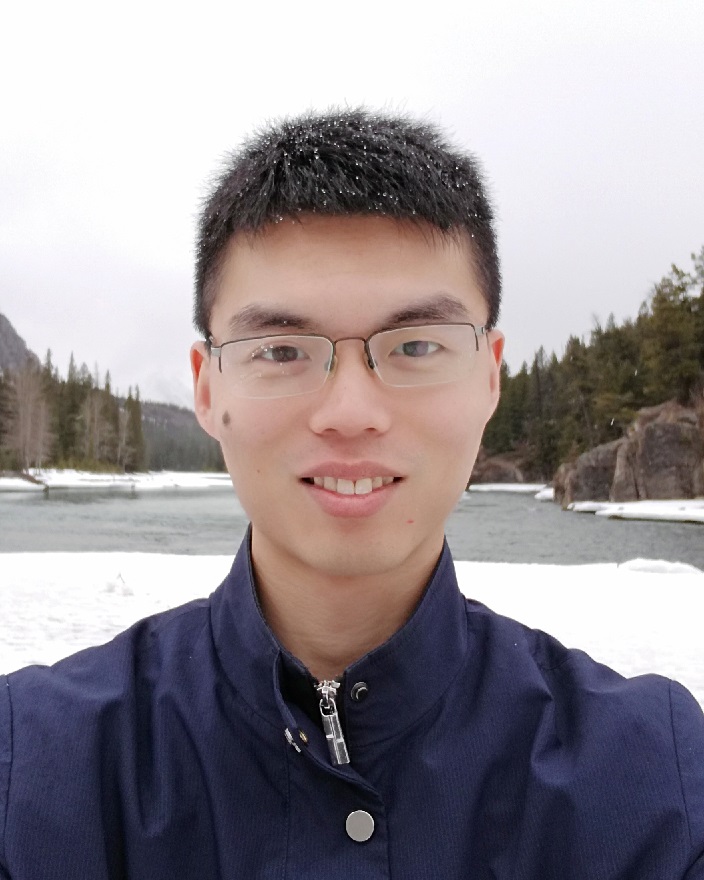}}]{Qiuqiang Kong}
(S'17) received the B.Sc. and M.E. degrees from South China University of Technology, Guangzhou, China, in 2012 and 2015, respectively. He is currently working toward the Ph.D. degree from the University of Surrey, Guildford, U.K on sound event detection. His research topic includes sound understanding, audio signal processing and machine learning. He was nominated as the postgraduate research student of the year in University of Surrey, 2019. 
\end{IEEEbiography}

\begin{IEEEbiography}[{\includegraphics[width=1in,height=1.25in,clip,keepaspectratio]{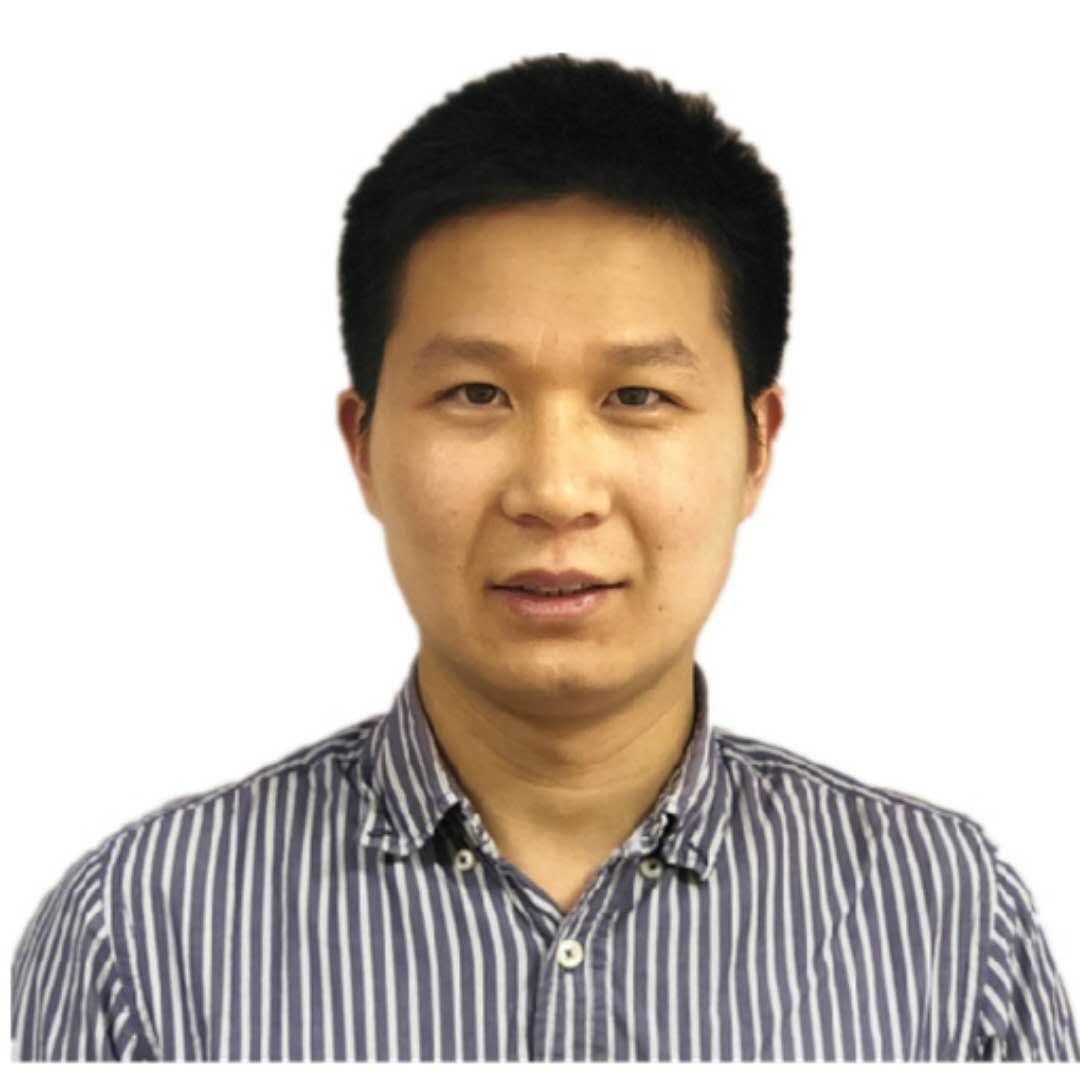}}]{Yong Xu}
(M'17) received the Ph.D. degree from the University of Science and Technology of China (USTC), Hefei, China, in 2015, on the topic of DNN-based speech enhancement and recognition. Currently, he is a senior research scientist in Tencent AI lab, Bellevue, USA.  He once worked at the University of Surrey, U.K. as a Research Fellow from 2016 to 2018 working on sound event detection. He visited Prof. Chin-Hui Lee's lab in Georgia Institute of Technology, USA from Sept. 2014 to May 2015. He once also worked in IFLYTEK company from 2015 to 2016 to develop far-field ASR technologies. His research interests include deep learning, speech enhancement and recognition, sound event detection, etc. He received 2018 IEEE SPS best paper award.
\end{IEEEbiography}

% if you will not have a photo at all:

% insert where needed to balance the two columns on the last page with
% biographies
%\newpage

\begin{IEEEbiography}[{\includegraphics[width=1in,height=1.25in,clip,keepaspectratio]{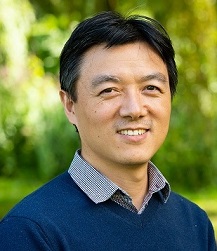}}]{Wenwu Wang}
(M'02-SM'11) was born in Anhui, China. He received the B.Sc. degree in 1997, the M.E. degree in 2000, and the Ph.D. degree in 2002, all from Harbin Engineering University, China. He then worked in King's College London, Cardiff University, Tao Group Ltd. (now Antix Labs Ltd.), and Creative Labs, before joining University of Surrey, UK, in May 2007, where he is currently a professor in signal processing and machine learning, and a Co-Director of the Machine Audition Lab within the Centre for Vision Speech and Signal Processing. He has been a Guest Professor at Qingdao University of Science and Technology, China, since 2018. His current research interests include blind signal processing, sparse signal processing, audio-visual signal processing, machine learning and perception, machine audition (listening), and statistical anomaly detection. He has (co)-authored over 200 publications in these areas. He served as an Associate Editor for IEEE TRANSACTIONS ON SIGNAL PROCESSING from 2014 to 2018. He currently serves as Senior Area Editor for IEEE Transactions on Signal Processing and Associate Editor for IEEE/ACM Transactions on Audio Speech and Language Processing.

\end{IEEEbiography}

\begin{IEEEbiography}[{\includegraphics[width=1in,height=1.25in,clip,keepaspectratio]{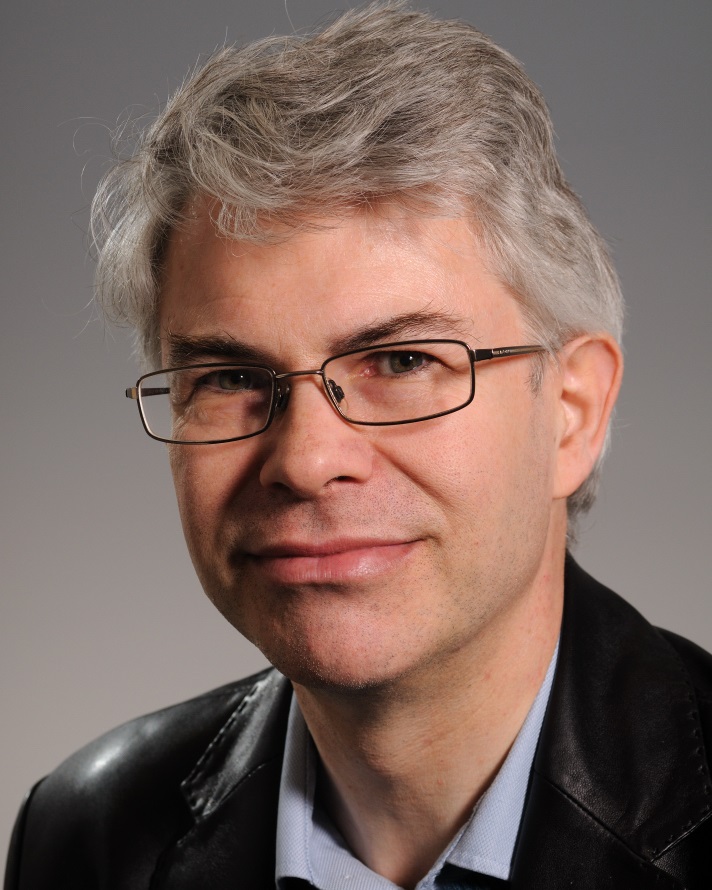}}]{Mark D. Plumbley}
(S'88-M'90-SM'12-F'15) received the B.A.(Hons.) degree in electrical sciences and the Ph.D. degree in neural networks from University of Cambridge, Cambridge, U.K., in 1984 and 1991, respectively. Following his PhD, he became a Lecturer at King's College London, before moving to Queen Mary University of London in 2002. He subsequently became Professor and Director of the Centre for Digital Music, before joining the University of Surrey in 2015 as Professor of Signal Processing. He is known for his work on analysis and processing of audio and music, using a wide range of signal processing techniques, including matrix factorization, sparse representations, and deep learning. He is a co-editor of the recent book on 
Computational Analysis of Sound Scenes and Events,
and Co-Chair of the recent DCASE 2018 Workshop on Detection and Classifications of Acoustic Scenes and Events. He is a Member of the IEEE Signal Processing Society Technical Committee on Signal Processing Theory and Methods, and a Fellow of the IET and IEEE.
\end{IEEEbiography}

% You can push biographies down or up by placing
% a \vfill before or after them. The appropriate
% use of \vfill depends on what kind of text is
% on the last page and whether or not the columns
% are being equalized.

%\vfill

% Can be used to pull up biographies so that the bottom of the last one
% is flush with the other column.
%\enlargethispage{-5in}

% that's all folks
\end{document}